\documentclass[11pt]{article}
\usepackage{mystyle}

\title{Twinac: A Universal Framework for Virtual Accelerator Controls\\{\footnotesize \texttt{FERMILAB-PUB-25-0483-AD-DI-PIP2}}\\\vspace{-0.4cm}{\footnotesize \texttt{ArXiv Draft}}}

\author{
  Tia Miceli, Abhishek Pathak, Aaron G Sauers\\
  Fermi National Accelerator Laboratory\\
  Batavia, IL 60510, USA\\
  \texttt{miceli@fnal.gov}, \texttt{abhishek@fnal.gov}, \texttt{asauers@fnal.gov}
}

\date{\today}

\begin{document}
\maketitle
\thispagestyle{fancy}

\begin{abstract}
We propose a universal framework for a system of virtual accelerator controls (Twinac), a standard toolkit for research institutions to design, maintain, and use a real-time, end-to-end ``digital twin'' of their particle accelerator facility.
This virtual counterpart will mirror any physical accelerator to provide (1) predictive maintenance; (2) surveillance of hidden environmental factors, such as seasonal temperature variations, which could impact performance of power supplies, magnets, and other instruments; and (3) a capability to model novel ways to operate the accelerator without risking equipment damage.
Twinac is envisioned as accelerator facility agnostic, allowing institutions to share and reuse myriad simulation approaches (analytics-based, A.I.-driven, or combinations of the like) across facilities.
The Twinac system lays the groundwork for a collaborative network of institutions to maintain and update this shared virtual accelerator technology.
\end{abstract}

\vspace{0.5cm}

\renewcommand{\thefootnote}{\fnsymbol{footnote}}

\section{Introduction and Background}

\subsection{Digital Twin Technology Overview}
A ``digital twin'' (DT) is a non-physical copy of a real-world system that closely mirrors its properties and behaviors in a digital format.
The concept was first introduced in 2002 for product lifecycle management~\cite{Grieves2016DigitalSystems} to improve how products are designed, manufactured, and maintained.
DTs have since been applied in diverse areas, including factories, assembly lines, power plants, wind turbines, hospitals, and commercial farms~\cite{Fuller2020DigitalResearch}.
They are used for predictive maintenance, prototyping, and testing controllers and regulators for their physical counterparts.

\begin{figure}[htbp]
  \centering
  \begin{minipage}[b]{0.48\textwidth} 
    \centering
    \includegraphics[width=\textwidth]{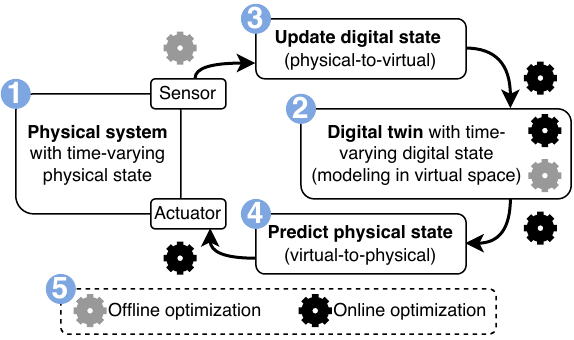}
    \caption{Adapted from~\cite{Thelen2022ATechnologies} the five components of a digital twin system.}\label{fig:5d-dt}
  \end{minipage}
  \hfill
  \begin{minipage}[b]{0.5\textwidth}
    \centering
    \includegraphics[width=\textwidth]{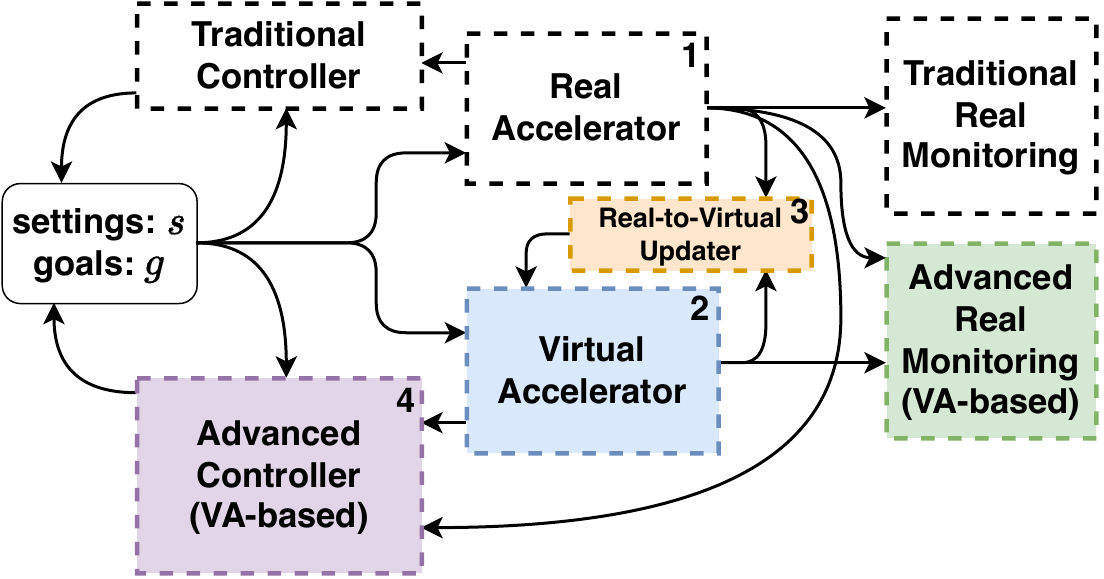}
    \caption{Sketch of how a digital twin system is applied to particle accelerators to enable enhanced monitoring and feedback controls.}\label{fig:simpleVAusage}
  \end{minipage}
\end{figure}

A DT system is composed of five key components, Figure~\ref{fig:5d-dt}: The Physical System, Digital Twin, Update Digital State, Predict Physical State, and Offline and Online Optimization Algorithms. 
\begin{enumerate}
    \item The Physical System is the actual asset that requires monitoring and controlling. 
    \item The DT is a virtual version of that asset that copies how it behaves over time. 
    \item The Update Digital State keeps the DT accurate by using real-world sensor data to continuously update it. 
    \item The Predict Physical State uses the DT to inspect the system and recommend actions. 
    \item Finally, the Offline and Online Optimization Algorithms provide the methodological foundation to keep the digital and physical systems working synchronistically, both in real-time and in scheduled or planned updates.
\end{enumerate}

\subsection{Digital Twins in Particle Accelerator Context}
A virtual accelerator (VA)\footnote[1]{In the accelerator community the term ``predictive virtual accelerator'' is preferred; in technology and manufacturing disciplines the term ``digital twin'' is preferred. Both terms refer to the same thing, this paper uses the term ``VA''.} plays an integral role in modern accelerator operations.
It serves as a baseline for monitoring accelerators and facilitates anomaly detection as part of predictive maintenance systems~\cite{Jain2022TheFNAL, Radaideh2022ProgressModulators}.
VAs can also identify correlated but uncontrollable environmental factors (such as seasonal temperature variations that affect power supplies, magnets, and RF cavities) which can be incorporated into updated models. 

\begin{figure}[htbp]
  \centering
    \includegraphics[width=\textwidth]{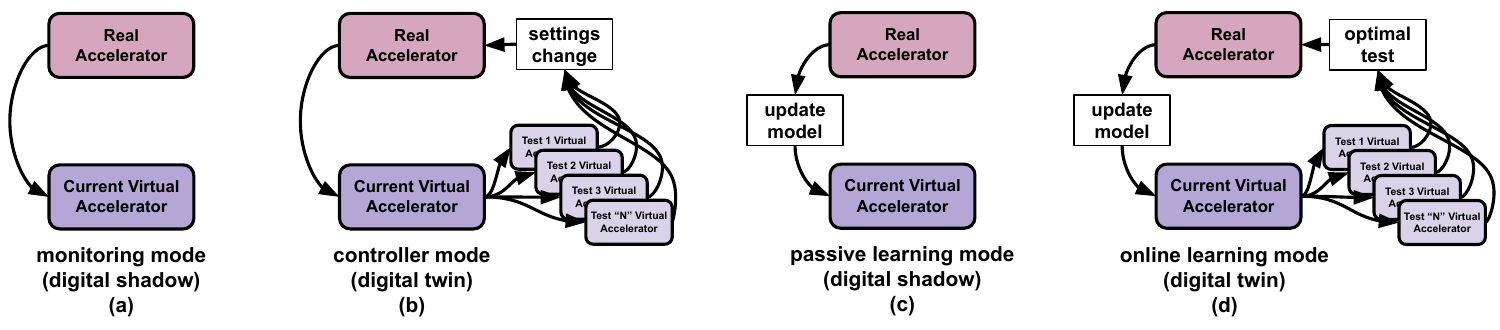}
    \caption{Modes for operating an accelerator digital twin (a) monitoring mode involves the virtual accelerator running in parallel with the real accelerator, automatically ingesting the same settings and comparing outputs to detect anomalies or performance drifts for predictive maintenance, (b) controlling, (c) passive learning, and (d) active learning.}\label{fig:DT_uses}
\end{figure}

Additionally, appropriately designed VAs, can provide the necessary integration interfaces with any accelerator control system.
Thus, allowing advanced tuning algorithms to be reused across facilities and control systems while offering a safe environment for testing new controllers. 
This reduces time needed for system commissioning. 
Figure~\ref{fig:DT_uses} illustrates the different modes of using and maintaining an accelerator DT for operations.

Interest in applying DT technology is growing throughout the accelerator community, demonstrating that the field is ready for a unified approach.
While several accelerator facilities have invested significant effort into developing their own facility-specific VA systems, the current landscape remains highly fragmented. 

Notable efforts have included the Bayesian optimization methodologies at SLAC~\cite{Roussel2023BayesianPhysics, Gupta2021ImprovingLearning} and Brookhaven National Laboratory's physics-based simulator to pre-train a reinforcement learning (RL) algorithm for controlling particle bunch merging at the Relativistic Heavy Ion Collider (RHIC)~\cite{Nguyen2024AComplex}.
Lawrence Berkeley National Laboratory has developed a novel adaptive machine learning algorithm that predicts the full 6-dimensional structure of electron bunches in HiRES, a compact particle accelerator used for ultra-fast electron diffraction~\cite{Scheinker2023AdaptiveAccelerator}.
In Europe, initial discussions have begun on a proposal to modernize the ``Matlab Middle Layer''~\cite{PortmannAControl} for light sources, by creating a Python-based Middle Layer (PML) to support VAs that work with EPICS and Tango control systems. 

However, many of these efforts suffer common limitations: They are often specific to individual facilities, depending on specific simulation codes, and lack standardized interfaces that would allow components to be reused across different systems. 

\subsection{The Need for a Universal Framework}
Accelerator facilities worldwide face common operational challenges that VA technology can help address, yet current efforts remain fragmented, creating significant barriers to collaboration and code reuse.
This fragmentation presents both a challenge and an opportunity for community-wide standardization that could accelerate innovation across the field.

The interdisciplinary nature of VA development spans at least five disciplines, emphasizing the need for a unified infrastructure that supports collaboration across roles. 

Accelerator and beam physics computational researchers can import their simulations for use in VA construction, fine-tuning, and operations. 
Subsystem and controls engineers can prototype controller algorithms in safe, virtual environments that closely mimic real accelerators, supported by libraries that allow reuse at other facilities.
Artificial intelligence and machine learning (AI/ML) researchers can contribute ultra-fast surrogate models of VA components and develop methods for combining different models.
Accelerator operations experts and data scientists can collaborate on algorithms for anomaly detection, predictive maintenance, and adaptations to changing external conditions.
Accelerator design researchers can use all available assets to prototype full end-to-end VAs prior to beginning large projects like a Higgs factory or muon collider.

A universal framework would allow for community-wide standardization, enabling
researchers and engineers to share models, algorithms, and best practices, thereby precipitating innovation at facilities throughout the field. The Twinac framework aims to break down disciplinary silos within institutions by supporting collaboration among accelerator and beam physics computational researchers, subsystem and controls engineers, AI/ML researchers, operations experts, and accelerator design researchers.

The Twinac framework described herein aligns with the HEP Program Office mission~\cite{DOEOfficeofScience2025}, the P5 2023 recommendations for software and computing~\cite{Murayama_2023}, and the GARD Program’s Accelerator and Beam Physics Roadmap~\cite{J.Blazey2023AcceleratorRoadmap}, which identifies VA systems as APB Grand Challenge 4.

\section{Twinac Framework Architecture and Design}
Twinac will be designed as a universal, facility-agnostic framework.
Figure~\ref{fig:simpleVAusage} shows how the settings and goals selected or approved by human operators are used in traditional accelerator control and monitoring (un-shaded boxes) and the necessary components to enable DT operations.
\begin{itemize}
    \item Universal Applicability: Will be agnostic to specific simulation codes, surrogate model\\methodologies, host OS, control systems, and particle beam types.
    \item Community Standards Adherence: Will strictly adhere to established standards like\\CAMPA~\cite{VayCAMPAAccelerators}, openPMD~\cite{Huebl2015OpenPMDData.}, and PICMI~\cite{PICMIcontributers2024Particle-In-CellDocumentation}.
    \item Modular Architecture: Will enable mix-and-match capabilities for VA components.
    \item Separation of Concerns: Will ensure interoperability by decoupling VA components from
control system interfaces~\cite{IngenoSoftwareArchitectsHandbook}.
\end{itemize}

\subsection{Fundamental Assumptions}
\begin{figure}[htbp]
  \centering
    \includegraphics[width=0.8\textwidth]{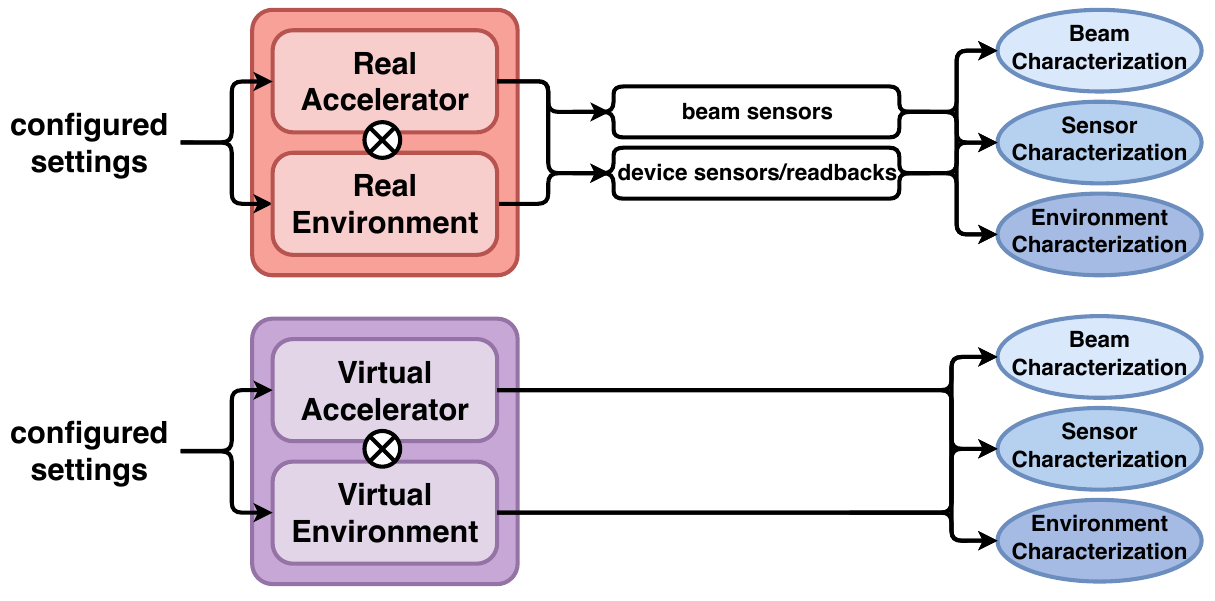}
    \caption{As configured settings are sent to the real accelerator (RA), data is collected by beam
sensors and facility readbacks from the RA and environment. From these data, the characterization of the beam sensors, and environment can be quantified. These characteristics can be compared with those produced by the virtual accelerator (VA) and virtual environment.}\label{fig:DT-DataFlow}
\end{figure}

Figure~\ref{fig:DT-DataFlow} shows the dataflow for the RA and the VA.
The RA interacts with the real environment to produce three fundamental types of data that are monitored: beam, sensors/devices, and environment.
When the VA has a high-fidelity, they should match those generated by the VA through its interactions with the virtual environment.

This separation of observables and predictions into beam, sensors/devices, and environment allows for more accurate modeling and monitoring of the factors that affect accelerator performance.
For example, if the beam characterization needs to be modified, the changes can be tested in the VA before being applied to the RA.
If the characterizations of all three factors match after applying the new settings, the VA is considered accurate.
If they do not match, the VA needs further improvement; see Section~\ref{sec:ops_suite} for more details.

\subsection{Twinac Software Suites}\label{sec:sw_components}
Twinac will have three software suites used in different phases of the DT lifecycle.
First, users will design and build a DT of an accelerator facility with the Twinac Designer Suite.
Next, the Twinac Commissioning Suite will tune the VA to the RA.
Finally, the Twinac Operations Suite will interface with the RA control system to use the VA for monitoring and control.

\subsubsection{Twinac Designer Suite}
The framework will facilitate the reuse of universal (Platonistic) accelerator component models, such as analytical simulations, AI/ML-based heuristic or surrogate models, and hybrid models,  and actions/decisions that are common across accelerator facilities.
For example, every facility has a dipole magnet, a vacuum pump, and an RF cavity.
Digital models of these VA elements can be shared and fine-tuned using the framework to match the corresponding RA components.
During commissioning, analytical simulations are tuned via modified parameters or scaling factors. 
For AI/ML-based heuristic models, tuning is accomplished using various transfer learning techniques to update the model weights.

\begin{figure}[htbp]
  \centering
    \includegraphics[width=0.9\textwidth]{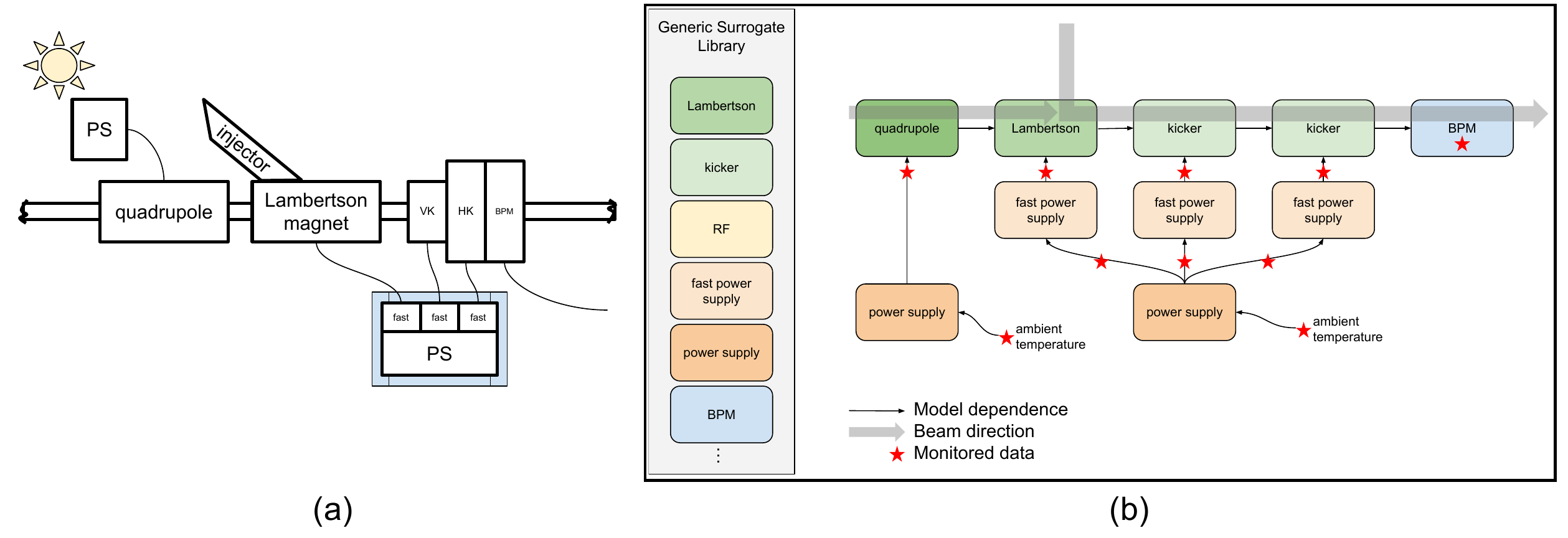}
    \caption{Example of a user GUI for developing a skeleton of a DT.}\label{fig:DT_builder}
\end{figure}

\begin{enumerate}
    \item \textbf{Designer GUI:} Will allow users to select base models (from simulation or surrogate libraries) to compose a DT (See Figure~\ref{fig:DT_builder}).
    \item \textbf{Virtual Accelerator (VA) Components Library:} This module will provide the building blocks for VAs, supporting multiple simulation codes (MAD-X~\cite{MAD-XCERN}, Synergia~\cite{SynergiaSynergia}, BLAST~\cite{OVERVIEWCODES}, TraceWin~\cite{TraceWinExplanation}), AI/ML-based surrogate models (GBMs, MLPs, VAEs), and hybrid approaches. It uses component abstraction for standardization. Each component category has a list of implementations from which to choose. Each component also has configurable settings and computes its effect on the beam.
    \item \textbf{Glue Library for Component Integration:} This module will provide methodologies for combining VA components serially and hierarchically, managing accumulative and non-linear effects to create realistic composite VA predictions.
    \item \textbf{VA Design File:} This will serve as the backend to the DT builder, storing user selections. This can be version-controlled, and instantiated multiple times to test VA responses in parallel.
\end{enumerate}

\subsubsection{Twinac Commissioning Suite}
\begin{figure}[htbp]
  \centering
    \includegraphics[width=\textwidth]{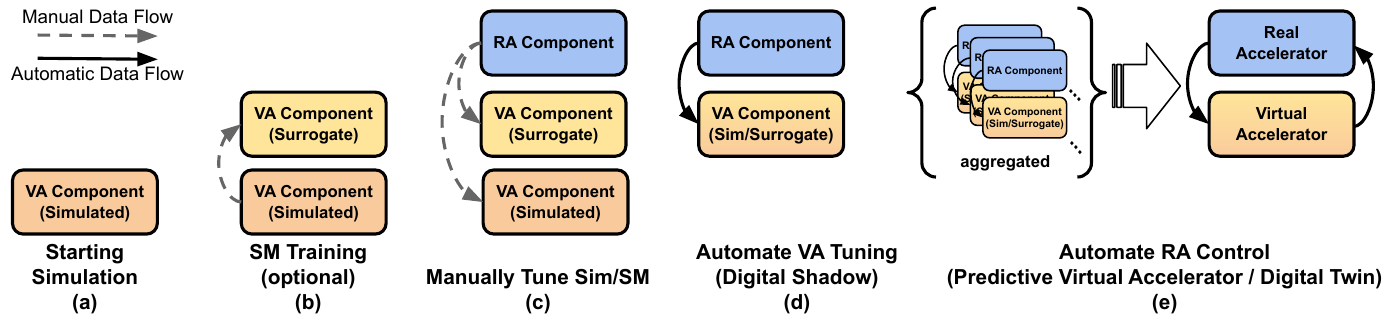}
    \caption{Commissioning of accelerator DT components from base simulation, or from data-driven surrogates. Surrogate model and simulation model are abbreviated SM and ``sim,'' respectively.}\label{fig:DT_commissioning}
\end{figure}

Figure~\ref{fig:DT_commissioning} shows the progressive steps for creating a virtual particle accelerator (Note that the VA component can either be embodied by a simulation or a heuristic surrogate model as defined in the previous section Twinac Design Suite). 
Dashed lines show manual data flow; solid lines show automated data flow.
Step (a) and (b) show an initial estimated virtual model based either on simulation, or surrogate models.
Step (c) shows the initial tuning of the VA to match RA data.
Step (d) shows a digital shadow automatically updated with the RA.
Shadows can be maintained per component, or per aggregates of components.
Step (e) shows a DT system, wherein the VA is used to make decisions in the RA, as in the next section.

Toward the end of commissioning, the modules (b) and (c) shown in Figure~\ref{fig:DT-Modes} are used to fine-tune the VA.
The R2V Tuner modifies VA parameters to match the RA.
The R2V Uncertainty Minimizer uses the VA to compute the settings necessitating configuration within the RA to minimize the VA prediction uncertainties.

\subsubsection{Twinac Operations Suite}\label{sec:ops_suite}
\begin{figure}[htbp]
  \centering
    \includegraphics[width=0.85\textwidth]{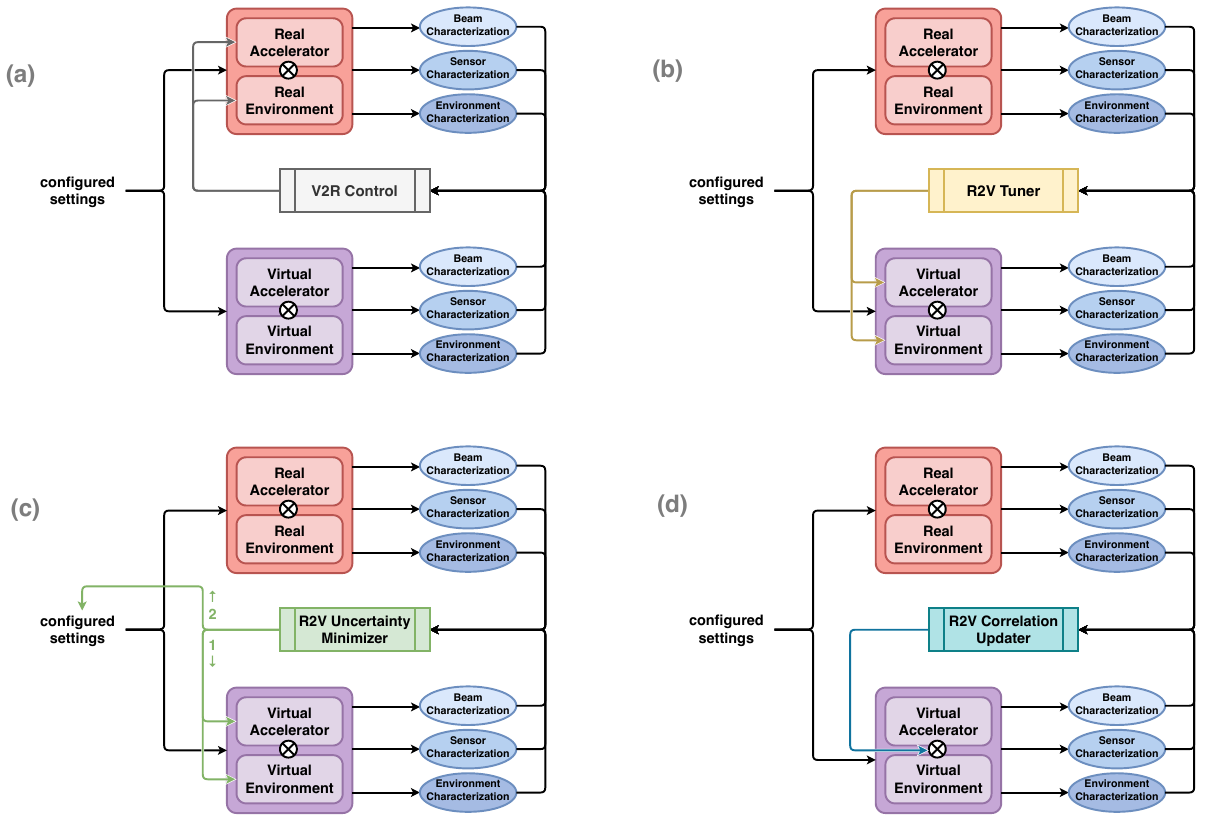}
    \caption{Modes in which the VA and RA interact. (a) Using VA tests to control RA. (b) Tuning VA based on current RA. (c) Fine-tuning VA based on current RA. (d) Tuning accelerator correlations with the environment. }\label{fig:DT-Modes}
\end{figure}

Four main \textit{action modules} are required for operations, as shown in Figure~\ref{fig:DT-Modes}.
\begin{enumerate}
    \item \textbf{V2R Control Module:} The ``Virtual-to-Real control'' module will test multiple settings virtually and choose new ones that will bring the RA to the desired state, see Figure~\ref{fig:DT-Modes}~(a). It will support integration with optimization algorithms, such as XOpt~\cite{Roussel2022XOPT:ALGORITHMS}, including both static controllers (i.e., traditional optimization) and dynamic controllers (reinforcement learning~\cite{SuttonIntroduction}, and adaptive ML~\cite{Scheinker2023AdaptiveAccelerator}), and offers support for multi-level autonomy~\cite{Huang2008Autonomy2.1}.

    \item \textbf{R2V Tuner Module:} The ``Real to Virtual Tuner'' will adjust the weights and model parameters to bring the VA into alignment with the RA, see Figure~\ref{fig:DT-Modes}~(b).

    \item \textbf{R2V Uncertainty Minimizer Module:} The ``Real-to-Virtual Uncertainty Minimizer'' will test multiple virtual settings and chooses those which will reduce the VA uncertainty to a desired level, then selects the best settings based on the updated VA, Figure~\ref{fig:DT-Modes}~(c). A collection of algorithms is available for this purpose.

    \item \textbf{R2V Correlation Updater Module:} The ``Real-to-Virtual Correlation Updater'' will add or modulate the effects of the environment within the VA so that the VA matches the RA, see Figure~\ref{fig:DT-Modes}~(d). This module will activate when the variance remains unsatisfactory after trying the R2V Tuner and the R2V Uncertainty Minimizer modules.
\end{enumerate}

\begin{figure}[htbp]
  \centering
    \includegraphics[width=0.75\textwidth]{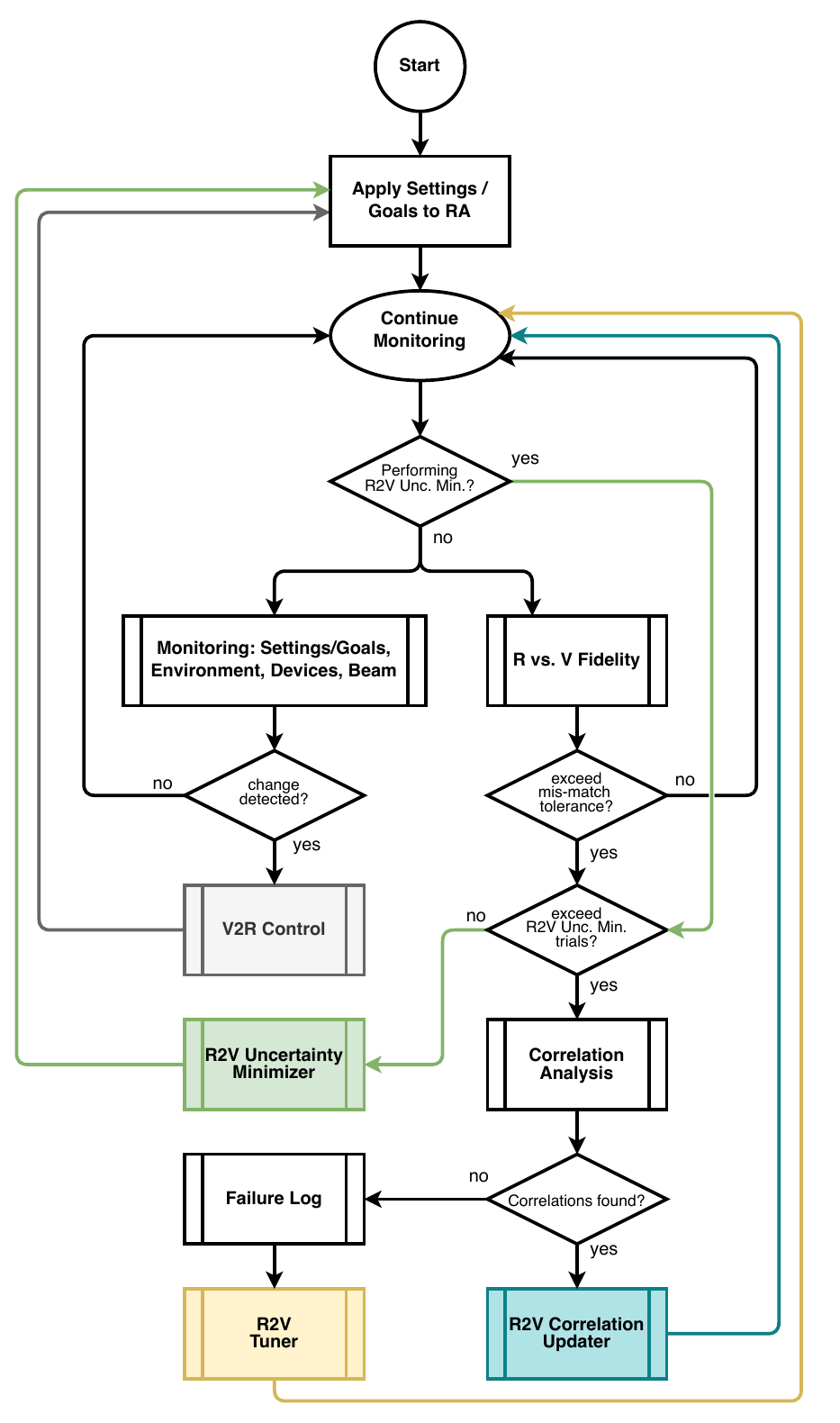}
    \caption{The overall flow for using a VA with an RA to provide control functionality, monitoring, and refinement of the VA. Flow described in Section~\ref{sec:ops_suite}.}\label{fig:DT-Operations}
\end{figure}

The operational flow using these action modules are outlined in the flowchart of Figure~\ref{fig:DT-Operations}.
It requires an additional set of four \textit{analysis modules}.

\begin{enumerate}
    \item \textbf{Monitoring Module:} This module will watch for changes in settings/objectives, environment, devices, and beam characteristics. As changes are desired, the V2R Control module will apply the appropriate changes. It will also watch for anomalies and failure modes to alert operators.
    
    \item \textbf{R vs. V Fidelity Module:} This module will evaluate how well the VA is representing the RA. Various algorithms and thresholds can be chosen to ensure that the user knows the reliability of the VA predictions.
    
    \item \textbf{Correlation Analysis Module:} This module will identify parameters that are correlated with the RA-VA discrepancy. These parameters will be passed to the R2V Correlation Updater module if found so that these new parameters or weights can be incorporated into the VA.

    \item \textbf{Failure Log Module:} If no new correlated parameters are identified, then the VA and the state of the RA will be recorded in a Failure Log.
\end{enumerate}

As the Failure Log grows, users will be able to analyze these offline and compare with any other databases (weather logs, building maintenance logs, train time-tables, moon phases, etc.) to discover ``unknown unknowns'' and hidden variables.
Users may iteratively update the supplemental environmental factors to improve the VA fidelity.

\subsection{Technology Stack and Implementation}
Further planning is needed to decide the most appropriate implementation language for each framework component, for example, RUST, Python, C++, Julia and others should be considered.
The design includes hardware abstraction for CPU, GPU, and FPGA support, and agnostic interfaces for control systems.
Containerization via Docker will ensure the code is portable and behaves identically across different environments.
These containerized workloads will be managed using Kubernetes~\cite{kubernetes}, the industry-standard open-source platform for container orchestration.

Kubernetes provides a resilient, scalable foundation for hosting a VA.
By hosting the Twinac in Kubernetes, the Monitoring Module can interface with a separate technology, the Machine Learning Operations (MLOps) pipeline~\cite{MLOpsDT}, being developed for automating the management of AI/ML models.
MLOps is a robust ecosystem that automates the process of training, validating, deploying, and monitoring the models that form the core of the VA.

\subsection{Licensing Framework for Interoperability}
The success of Twinac as a universal framework depends on establishing clear licensing practices that enable collaboration across institutions while respecting intellectual property requirements.
To accomplish this, we propose a licensing approach that draws from best practices in research software development, encouraging openness, but also allowing flexibility.

The framework adopts the permissive Apache 2.0 software license~\cite{apache_license_2_0}, which supports collaboration among multiple organizations.
One key benefit of Apache 2.0 is its explicit patent grant that protects users from being sued by those who contributed code.
This patent safeguard extends to anyone who uses the software, creating a conducive environment for innovation.

Individual VA components, such as simulations, surrogate models, or control algorithms, may carry their own licenses, chosen by the institutions that created them, so long as they remain compatible with Twinac's interfaces.
To ensure legal clarity and protect all participants, Twinac can use a well-tested Contributor License Agreement (CLA) template~\cite{apache_contributor_agreements} from the Apache Software Foundation.
This CLA lets contributors keep ownership of their code while allowing the consortium to reuse and share it, creating a clear record of provenance.

For aspects of the software that need to stay proprietary, the framework allows a plugin system:  Research institutions can connect closed source modules to Twinac through well-defined APIs, without having to share their source code.
This helps protect sensitive or export-controlled technology while still encouraging broad participation. 
The key requirement is that interfaces remain open and documented, ensuring that alternative implementations and individual VA components can be developed by the community.

This proposed framework includes license compatibility verification to ensure that software components with different licenses can legally be combined and distributed together. 
Because Twinac establishes compatibility requirements at project inception, developers can make informed decisions from the beginning, rather than run into legal issues later. 
This allows teams to select compatible dependencies and design architectures that meet licensing requirements from the start, rather than to discover conflicts during integration with Twinac libraries. 
The Apache 2.0 license’s compatibility with both GPL version 3~\cite{gplv3} and proprietary software makes it particularly suitable for the diverse ecosystem of accelerator software. 

By establishing clear licensing practices from the outset, the community can build a flexible, unified framework rather than attempt to reconcile problems after years of incompatible development.

\section{Enabling Efficient Research}
The open and shared nature of the proposed Twinac infrastructure and libraries creates a plug-and-play environment testing competing versions of VA components.
VA components developed at one accelerator facility, should be readily testable at another facility.
The most prolific VA research areas that can be easily compared and benchmarked are (1) surrogate models, (2) methods for integrating successive and/or hierarchical surrogate models and simulations, and (3) control algorithms for automation. 

\subsection{Opportunities in Surrogate Model Benchmarking}
For any physical process, a surrogate model can serve as a shortcut version for a complex simulation, which may be too resource-heavy for timely predictions.
Examples of physical processes include how high space-charge beams are accelerated and focused and non-linear effects from magnetic fringe fields.
Examples of AI/ML tools that could be used for this are GBMs, MLPs, and VAEs.
These can be combined with state-of-the-art uncertainty quantification (UQ) techniques~\cite{Thelen2023APerspectives,Abdar2021AChallenges} to estimate the reliability of model predictions.

For each accelerator use-case, the Twinac framework will streamline testing and benchmarking models side by side.
Users will be able to compare their accuracy, mean squared error, and computational performance
For example, the framework could look at how long it takes a model to make a prediction (inference time), and how long it takes to train it (training time, a critical factor if the system needs frequent updates).

\subsection{Opportunities in Benchmarking of Model Aggregation Methodologies}

A key challenge and opportunity in modeling complex systems is figuring out how to connect diverse simulations and surrogate models for different components. 
Twinac’s proposed use of shared formats for describing particle beams  will be the first step in ensuring surrogate models can work with established high-fidelity simulation codes like MAD-X~\cite{MAD-XCERN} and BLAST~\cite{OVERVIEWCODES}.
Just as Twinac will simplify testing individual models, it will also make it easy to test various aggregation methods, which will expedite the process.

As the accelerator community works on linking digital models, Twinac should first consider using existing open-source tools that already  focus on these integration challenges. 
One tool under investigation is the Surrogate Modeling Toolbox (SMT) 2.0~\cite{Saves2024SMTProcesses}, which has advanced features for handling hierarchical models, wherein one model depends on another.
These features are a good fit for the modeling accelerators, which are complex and involve multivariate systems.

Digital models can be linked in different ways: serially, hierarchically, or even by combining their outputs.
For example, the output of one model could be: fed directly into the input of another; adjusted using a scaling factor or function; or connected via a small artificial neural network.

Each of these connection strategies should be systematically tested, using different combinations of models, such as surrogate-to-surrogate, sim-to-sim, and hybrid mixes.
Twinac's framework will streamline the comparison their accuracy and performance against high-fidelity simulations.

\subsection{Opportunities in Control Algorithm Benchmarking}
The Twinac framework will allow researchers to test and validate different control algorithms on the DT before applying them to the physical accelerator.
For example, static optimization algorithms (e.g., XOpt~\cite{Roussel2022XOPT:ALGORITHMS}) can run simulations to find the ideal control settings to try next in order to determine the optimal setpoint.

Likewise, dynamic control methods, such as Reinforcement Learning (RL)~\cite{SuttonIntroduction} and Adaptive Machine Learning (AML)~\cite{Scheinker2023AdaptiveAccelerator} with in-situ policy-updating mechanisms, also can be tested safely on copies of the DT.
Twinac will make it easy to test different algorithms and compare how fast they find the optimal setting, how efficiently they explore options, and how well they handle growing numbers of parameters and objectives.

Once tested, these DT-based control optimizers can be used to safely connect the DT and physical accelerator to automatically carry out control and regulation tasks.

\section{Conclusions and Future Directions}
This work presents Twinac, a proposed comprehensive and community-driven framework for developing, tuning, and using universal VAs.
Twinac will be built on solid technical foundations and follow a practical, step-by-step implementation strategy to meet the growing demand for standardized VA tools across the global accelerator physics community.

The motivation for developing this framework is threefold: (1) to improve efficiency at real accelerator facilities, (2) to provide a safe and flexible environment for testing novel accelerator concepts, and (3) to facilitate the sharing of validated VA components among different facilities.

Looking ahead, key steps for launching and sustaining the Twinac collaboration include establishing a governing consortium, developing robust funding proposals for long-term support, integrating with next-generation accelerator initiatives, and expanding Twinac’s compatibility with additional accelerator types. 
By working together, the community can build a strong platform that benefits all participants and advances accelerator physics research and operations.

The success of this project depends critically on broad community participation and collaboration.
We therefore invite institutions worldwide to join us — by helping to build the core framework, creating control system interfaces and simulation codes, developing surrogate models and optimization tools, honing anomaly detection capabilities, testing hierarchical modeling approaches, and expanding software libraries.
Through this collective effort, we can establish a truly universal platform that meets the needs of the global accelerator community, and advances discovery and innovation.

\section*{Acknowledgments} 
This manuscript has been authored by Fermi Forward Discovery Group, LLC under Contract No. 89243024CSC000002 with the U.S. Department of Energy, Office of Science, High Energy Physics. Special thanks to \textbf{Jan Strube} for many conversations.

\FloatBarrier
\printbibliography

\end{document}